\documentstyle[twocolumn,pre,aps,epsfig]{revtex}

\begin{document}


\title{Symmetry breaking and coarsening in spatially distributed
evolutionary processes including sexual reproduction and disruptive
selection}

\author{Hiroki Sayama$^1$, Les Kaufman$^{1,2}$ and Yaneer Bar-Yam$^1$}

\address{$^1$ New England Complex Systems Institute, Cambridge, MA 02138\\
$^2$ Boston University, Dept.\ of Biology, Boston, MA 02215}

\maketitle

\begin{abstract} Sexual reproduction presents significant challenges to
formal treatment of evolutionary processes. A starting point for
systematic treatments of ecological and evolutionary phenomena has
been provided by the gene centered view of evolution which assigns
effective fitness to each allele instead of each organism. The gene
centered view can be formalized as a dynamic mean field approximation
applied to genes in reproduction / selection dynamics. We show that
the gene centered view breaks down for symmetry breaking and pattern
formation within a population; and show that spatial distributions of
organisms with local mating neighborhoods in the presence of
disruptive selection give rise to such symmetry breaking and pattern
formation in the genetic composition of local populations. Global
dynamics follows conventional coarsening of systems with nonconserved
order parameters. The results have significant implications for the
ecology of genetic diversity and species formation.
\end{abstract}


The dynamics of evolution can be studied by statistical models that
reflect properties of general models of the statistical dynamics of 
interacting systems\cite{baryam97}. Research on this topic can affect the
conceptual foundations of evolutionary biology, and many applications
in ecology, population biology, and conservation biology. Among the
central problems is understanding the creation, persistence, and
disappearance of genetic diversity. In this paper, we describe a model
of sexual reproduction which illustrates mean field approaches (the 
gene-centered view of evolution) and the relevance of symmetry breaking 
and pattern formation in spatially distributed populations as an example 
of the breakdown of these approximations.

Pattern formation in genomic space has been of increasing interest in
theoretical studies of sympatric
speciation\cite{baryam97,higgs91,bagnoli97,drossel00,kondrashov99,dieckmann99,higashi99}.
These papers advance our understanding of the mechanisms of forming
two species from one. However, they do not address the fundamental and
practical problems of genetic diversity and spatial inhomogeneity
within one species---a population whose evolution continues to be
coupled by sexual reproduction. Moreover, and significantly, these
papers do not address the implication of symmetry breaking and pattern
formation for the gene centered view as a fundamental framework of
evolutionary theory. In the following, we demonstrate that symmetry
breaking and pattern formation invalidate the gene centered view
(whether or not speciation occurs), and that they are important for
the spatio-temporal behavior of the genetic composition of sexually
reproducing populations. This has a wide range of implications for
ecology, conservation biology, and evolutionary theory.

Before introducing the complications of sexual reproduction, we start 
with the simplest iterative model of exponential growth of asexually 
reproducing populations:

\begin{equation}
N_i(t+1)=\lambda_i N_i(t)
\end{equation}

\noindent where $N_i$ is the population of type $i$ and $\lambda_i$ is
their fitness. If the total population is considered to be normalized,
the relevant dynamics is only of the proportion of each type, then we
obtain

\begin{equation}
P_i(t+1)=\frac{\lambda_i}{\sum_{i}\lambda_i P_i(t)} P_i(t)
\label{eqlabel2}
\end{equation}

\noindent where $P_i$ is the proportion of type $i$. The addition of
mutations to the model, $N_i(t+1)=\sum_j \lambda_{ij}N_j(t)$, gives
rise to the quasi-species model\cite{eigen89} which has attracted
significant attention in the physics community. Recent research has
focused on such questions as determining the rate of environmental
change which can be followed by evolutionary change.

Sexual reproduction causes offspring to depend on the genetic makeup
of two parents. This leads to conceptual problems (not just
mathematical problems) in evolutionary theory because the offspring of
an organism may be as different from the parent as organisms it is
competing against. A partial solution to this problem is recognizing
that it is sufficient for offspring traits to be correlated to
parental traits for the principles of evolution to apply. However, the
gene centered view is a simpler perspective in which the genes serve
as indivisible units that are preserved from generation to
generation\cite{footnote1}. In effect, different versions of the gene,
i.e.\ alleles, compete rather than organisms. This view simplifies the
interplay of selection and heredity in sexually reproducing organisms.

We will show, formally, that the gene centered view corresponds to a
mean field approximation\cite{baryam00}. This clarifies the domain of
its applicability and the conditions in which it should not be applied
to understanding evolutionary processes in real biological systems.
We will then describe the breakdown of the gene centered view in the
case of symmetry breaking and pattern formation and its implications
for the study of ecological systems.

It is helpful to explain the gene centered view using the ``rowers
analogy'' introduced by Dawkins\cite{dawkins89}. In this analogy boats
of mixed English- and German-speaking rowers are filled from a common
rower pool. Boats compete in heats and it is assumed that a speed
advantage exists for boats with more same-language rowers. The
successful rowers are then returned to the rower pool for the next
round. Over time, a predominantly and then totally same language rower
pool will result.  Thus, the selection of boats serves, in effect, to
select rowers who therefore may be considered to be competing against
each other\cite{footnote2}. In order to make the competition between
rowers precise, an effective fitness can be assigned to a rower. We
will make explicit the rowers model (in the context of genes and
sexual reproduction) and demonstrate the assignment of fitness to
rowers (genes).

The rowers analogy can be directly realized by considering genes with
selection in favor of a particular combination of alleles on genes. 
Specifically, for two genes, after selection, when allele $A_1$
appears in one gene, allele $B_1$ must appear on the second gene, and
when allele $A_{-1}$ appears on the first gene allele $B_{-1}$ must
appear on the second gene. We can write these high fitness organisms
with the notation $(1,1)$ and $(-1,-1)$, and the organisms with lower
fitness as $(1,-1)$ and $(-1,1)$. For simplicity, we assume below that
the lower fitness organisms are non-reproducing. Models which allow
them to reproduce, but with lower probabilities than the high fitness
organisms, give similar results.

The assumption of placing rowers into the rower pool and taking them
out at random is equivalent to assuming that there are no correlations
in reproduction (i.e. no correlations in mate pairing) and that there
is a sufficiently dense sampling of genomic combinations by the
population (in this case only a few possibilities). Then the offspring
genetic makeup can be written as a product of the probability of each
allele in the parent population. This assumption describes a
``panmictic population'' which forms the core of the gene centered
view often used in population biology. The assumption that the
offspring genotype frequencies can be written as a product of the
parent allele frequencies is a dynamic form of the usual mean field
approximation neglect of correlations in interacting statistical
systems\cite{footnote3}. While the explicit dynamics of this system is
not like the usual treatment of mean-field theory, e.g. in the Ising
model, many of the implications are analogous.

In our case, the reproducing parents (either $(1,1)$ or $(-1,-1)$)
must contain the same proportion of the correlated alleles ($A_1$ and
$B_1$) so that $p(t)$ can represent the proportion of either $A_1$ or
$B_1$ and $1-p(t)$ can represent the proportion of either $A_{-1}$ or
$B_{-1}$. The reproduction equation specifying the offspring (before
selection) for the gene pool model are:

\begin{eqnarray}
P_{1,1}(t+1) &=& p(t)^2 \\
P_{1,-1}(t+1) &=& P_{-1,1}(t+1) = p(t)(1-p(t)) \\
P_{-1,-1}(t+1) &=& (1-p(t))^2
\end{eqnarray}

\noindent where $P_{1,1}$ is the proportion of $(1,1)$ among the
offspring, and similarly for the other cases.

The proportion of the alleles in generation $t+1$ is given by the
selected organisms. Since the less fit organisms $(1,-1)$ and $(-1,1)$
do not survive this is given by
$p(t+1) = P'_{1,1}(t+1) + P'_{1,-1}(t+1)=P'_{1,1}(t+1)$, where
primes indicate the proportion of the selected organisms. Thus

\begin{equation}
p(t+1)=\frac{P_{1,1}(t+1)}{P_{1,1}(t+1)+P_{-1,-1}(t+1)}
\end{equation}

\noindent This gives the update equation:

\begin{equation}
p(t+1) = \frac{p(t)^2}{p(t)^2 + (1-p(t))^2} \label{eqlabel4}
\end{equation}

There are two stable states of the population with all organisms
$(1,1)$ or all organisms $(-1,-1)$. If we start with exactly 50\% of
each allele, then there is an unstable steady state. In every
generation 50\% of the organisms reproduce and 50\% do not. Any small
bias in the proportion of one or the other will cause there to be
progressively more of one type over the other, and the population will
eventually have only one set of alleles. This problem is reminiscent
of an Ising ferromagnet at low temperature: A statistically biased
initial condition leads to alignment.

This model can be reinterpreted by assigning a mean fitness (analogous
to a mean field) to each allele as in Eq.\ (\ref{eqlabel2}). The
fitness coefficient for allele $A_1$ or $B_1$ is $\lambda_1 = p(t)$
with the corresponding $\lambda_{-1} = 1-\lambda_1$. The assignment of
a fitness to an allele reflects the gene centered view.  The explicit
dependence on the population composition (an Engligh-speaking rower in
a predominantly English-speaking rower pool has higher fitness than
one in a predominantly German-speaking rower pool) has been objected
to on grounds of biological appropriateness \cite{sober82}. For our
purposes, we recognize this dependence as the natural outcome of a
mean field approximation.

We can describe more specifically the relationship between this
picture and the mean field approximation by recognizing that the
assumptions of no correlations in reproduction, a random mating
pattern of parents, is the same as a long-range interaction in an
Ising model. If there is a spatial distribution of organisms with
mating correlated by spatial location and fluctuations so that the
starting population has more of the alleles represented by 1 in one
region and more of the alleles represented by $-1$ in another region,
then patches of organisms that have predominantly $(1,1)$ or $(-1,-1)$
form after several generations. This symmetry breaking, like in a
ferromagnet, is the usual breakdown of the mean field approximation. 
Here, it creates correlations / patterns in the genetic makeup of the
population. When correlations become significant then the species has
two types, though they are still able to cross-mate and are doing so
at the boundaries of the patches. Thus the gene centered view breaks
down when multiple organism types form.

Understanding the spatial distribution of organism genotype is a
central problem in ecology and conservation
biology\cite{tilman97,durett94}.  The spatial patterns that can arise
from spontaneous symmetry breaking through sexual reproduction, as
implied by the analogy with other models, may be relevant.  A
systematic study of the relevance of symmetry breaking to ecological
systems begins from a study of spatially distributed versions of the
model just described. This model is a simplest model of disruptive
selection, which corresponds to selection in favor of two genotypes
whose hybrids are less viable. Assuming overlapping local reproduction
neighborhoods, called demes, the relevant equations are:

\begin{eqnarray}
p(x,t+1) &=& D(\bar{p}(x,t)) \\
D(p) &=& \frac{p^2}{p^2 + (1-p)^2}\\
\bar{p}(x,t) &=& \frac{1}{N_R}\sum_{|x_j|\le R} p(x+x_j,t) \\
N_R &=& \bigl| \{x_j \bigm| |x_j|\le R \} \bigr|
\end{eqnarray}

\noindent where the organisms are distributed over a two-dimensional
grid and the local genotype averaging is performed over a preselected
range of grid cells around the central cell. Under these conditions
the organisms locally tend to assume one or the other type. In
contrast to conventional insights in ecology and population biology,
there is no need for either complete separation of organisms or
environmental variations to lead to spatially varying genotypes.
However, because the organisms are not physically isolated from each
other, the boundaries between neighboring domains will move, and the
domains will follow conventional coarsening behavior for systems with
non-conserved order parameters.

\begin{figure}
\psfig{file=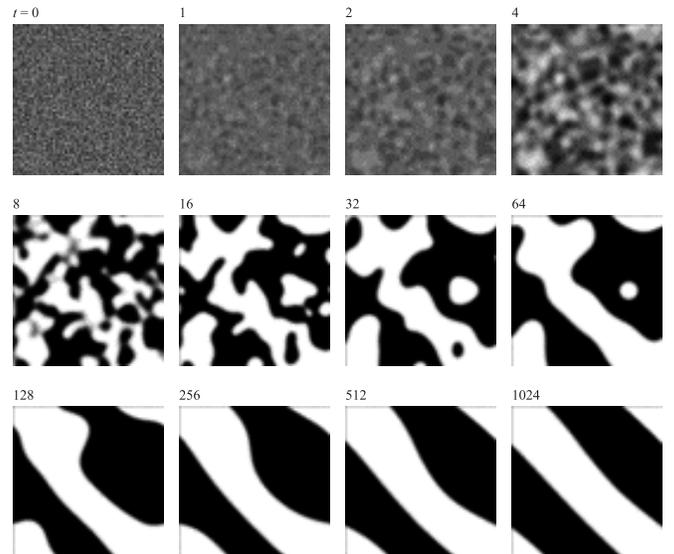,width=\columnwidth}
\caption{Spatially distributed evolution with disruptive selection
giving rise to two types appearing in patches and coarsening. The
space is periodic and has $256 \times 256$ sites, and the mating
neighborhood radius is $R=5$.}
\label{fig1}
\end{figure}

\begin{figure}
\psfig{file=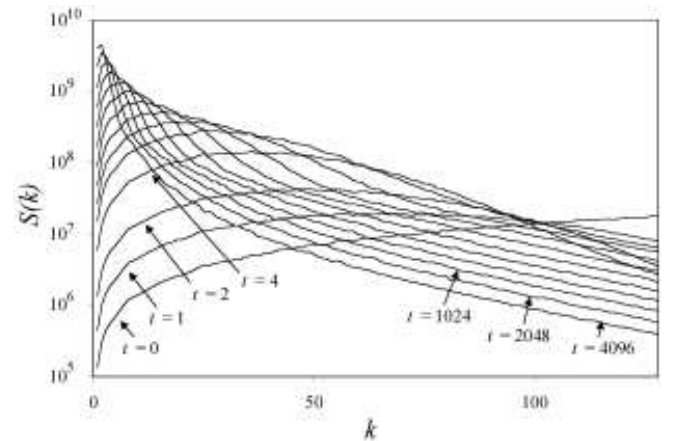,width=\columnwidth}
\caption{Fourier power spectra averaged over ten simulations of
evolutionary processes like that shown in Fig.\ \ref{fig1} ($512
\times 512$ sites and $R=1$).}
\label{fig2}
\end{figure}

\begin{figure}
\psfig{file=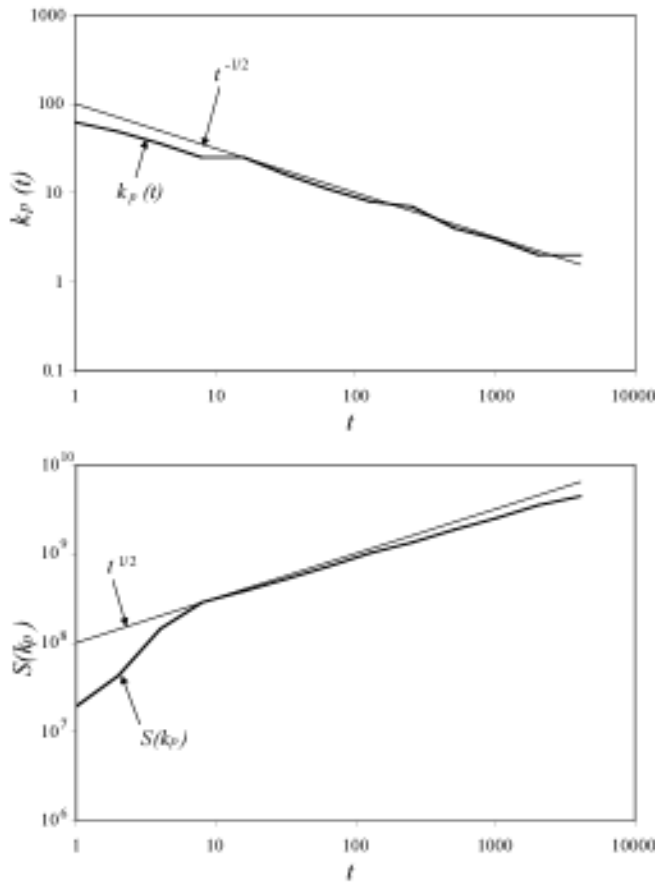,width=\columnwidth}
\caption{Temporal behavior of the peak of a Fourier power spectrum in
the shown case. Top: The peak frequency $k_p(t)$ which follows
approximately $t^{-1/2}$. Bottom: The peak power $S(k_p)$ which
follows approximately $t^{1/2}$.}
\label{fig3}
\end{figure}

\begin{figure}
\psfig{file=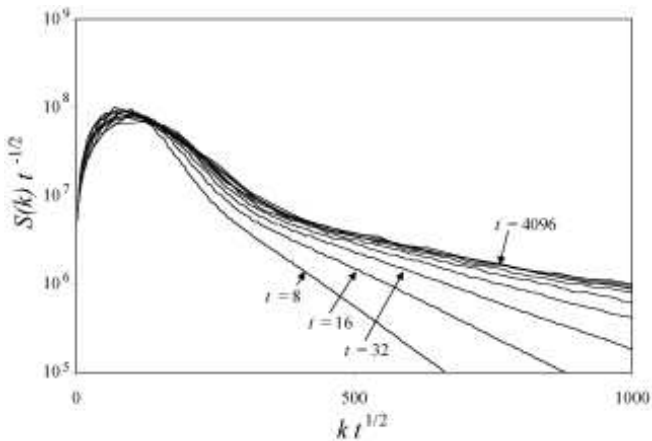,width=\columnwidth}
\caption{Collapsed version of the Fourier power spectra demonstrating
the scaling form $S(k)=t^{1/2} f(kt^{1/2})$.}
\label{fig4}
\end{figure}

A simulation of this model starting from random initial conditions is
shown in Fig.\ \ref{fig1}. This initial condition can arise when
selection becomes disruptive after being non-disruptive due to
environmental change. The formation of domains of the two different
types that progressively coarsen over time can be seen. While the
evolutionary dynamics describing the local process of organism
selection is different, the spatial dynamics of domains is equivalent
to the process of coarsening / pattern formation that occurs in many
other systems such as an Ising model or similar cellular automata
models\cite{bray94,ermentrout93}. Fourier transformed power spectra
(Figs.\ \ref{fig2}--\ref{fig4}) confirm the correspondence to
conventional coarsening by showing that the correlation length grows
as $t^{1/2}$ after initial transients. In a finite sized system, it is
possible for one type to completely eliminate the other type. However,
the time scale over which this takes place is much longer than the
results assuming complete reproductive mixing, i.e.\ the mean field
approximation.  Since flat boundaries do not move except by random
perturbations, a non-uniform final state is possible. The addition of
noise will cause slow relaxation of flat boundaries but they can also
be trapped by quenched (frozen) inhomogeneity.

The results have significant implications for ecology of genetic
diversity and species formation. The observation of harlequin
distribution patterns of sister forms is generally attributed to
nonhomogeneities in the environment, i.e. that these patterns reflect
features of the underlying habitat (=selective) template. Our
results show that disruptive selection can give rise to
spontaneously self-organized patterns of spatial distribution that
are independent of underlying habitat structure. At a particular time,
the history of introduction of disruptive selection events would
be apparent as a set of overlapping patterns of genetic diversity
that exist on various spatial scales.

\begin{figure}
\psfig{file=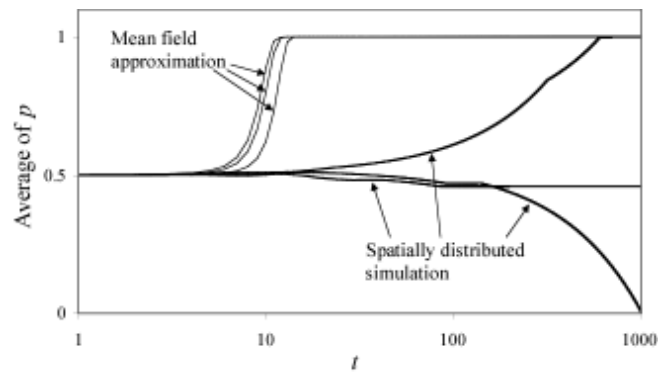,width=\columnwidth}
\caption{Comparison of the time dependence of type probability in the
mean field approximation and symmetry breaking, calculated using different
random number sequences. Diversity persists much longer in the latter. In
some cases, forever.}
\label{fig5}
\end{figure}

More specific relevance of these results to the theoretical
understanding of genetic diversity can be seen in Fig.\ \ref{fig5}
where the population averaged time dependence of $p$ is shown. The
gene centered view / mean field approximation predicts a rapid
homogenization over the entire population.  The persistence of
diversity in simulations with symmetry breaking, as compared to its
disappearance in mean field approximation, is significant.
Implications for experimental tests and methods are also important.
Symmetry breaking predicts that when population diversity is measured
locally, rapid homogenization similar to the mean field prediction
will apply, while when they are measured over areas significantly
larger than the expected range of reproduction, extended persistence
of diversity should be observed.

The divergence of population traits in space studied in our work can
also couple to processes of speciation, i.e., processes that prevent
interbreeding or doom the progeny of such breedings. These may include
assortative mating, whereby organism traits inhibit interbreeding. Such
divergences can potentially lead to the formation of multiple species
from a single connected species (sympatric speciation). By contrast,
allopatric speciation, where disconnected populations diverge, has
traditionally been the more accepted process even though experimental
observations suggest sympatric speciation is important.

Recent
studies\cite{baryam97,higgs91,bagnoli97,drossel00,kondrashov99,dieckmann99,higashi99}
have begun to connect the process of symmetry breaking to sympatric
speciation. Without considering pattern formation in physical space,
we and other researchers have been investigating the role of pattern
formation in genomic space as a mechanism or description of sympatric
speciation. These studies include: a model of stochastic branching and
fixation of subpopulations due to genetic drifts and local
reproduction in genome space\cite{higgs91}, general reaction-diffusion
Turing pattern formation models in genomic
space\cite{baryam97,bagnoli97,drossel00}, and specific
individual-based models of reproductive isolation involving
assortative mating and disruptive selection (intrinsic disruptive
selection, or disruptive selection arising from competition or sexual
selection)\cite{kondrashov99,dieckmann99,higashi99}. Our work,
presented here, is unique in discussing spatial inhomogeneity and
genetic diversity within one species.

In conclusion, in formalizing sexual reproduction in evolutionary
theory, we have found fundamental justification for rejecting the
widespread application of the gene centered view. The formal
mathematical analysis we presented to demonstrate the lack of
applicability of the gene centered view is an essential step toward
developing a sound conceptual foundation for evolution. We also
showed that the gene centered view breaks down for species where local
mating and disruptive selection give rise to symmetry breaking and
pattern formation, which correspond to genetic inhomogeneity and trait
divergence of subpopulations. The patterns formed undergo coarsening,
following the usual universal spatio-temporal scaling behavior. The
slow movement of boundaries between types cause long term persistence
of genetic diversity through the local survival of (partially)
incompatible types. This provides a new understanding of the
development and persistence of spatio-temporal patterns of genetic
diversity within a single species.

One should note that the context in which the gene centered view
breaks down is of profound significance in applied aspects of modern
ecology and conservation biology. The preservation of endangered
species and ecosystems is currently at risk due to a dramatic decrease
in their genetic diversity. We have described the implications of our
results for the experimental observation of genetic diversity in
endangered species. Our study of spatial patterns of genetic diversity
in populations may also help guide the design of conservation areas
and human directed breeding programs for endangered organisms.


\begin{references}
\bibitem{footnote1} The evolutionary indivisibility of genes is also
not necessarily well justified.
\bibitem{footnote3} The general relationship between the
mean-field approximation and the gene centered view can be shown by
considering a two-step process of reproduction and selection:
\begin{eqnarray*}
\{N(s;t)\} & = & R[\{N'(s;t-1)\}] \\
\{N'(s;t)\} & = & D[\{N(s;t)\}]
\end{eqnarray*}
\noindent where $s$ is a particular genome, and $N, N'$ are numbers of
reproduced, selected organisms respectively. Using a mean field
approximation for offspring, the probability of
a particular genome $s$ is the
product of probabilities of the alleles $a_i$: $P(a_1,\ldots,a_N) =
\prod P(a_i)$. This enables the two-step
update equations to be written as a one-step update equation for each
allele:
\[n'(a_i;t) = \tilde{D}[\{n'(a_i;t-1)\}] = \lambda(\{n'\}) n'(a_i;t-1)\]
\noindent where $n'(a_i)$ is the number of allele $a_i$. For details
see \cite{baryam00}.
\end{references}
\end{document}